\documentclass[12pt]{article}

 \usepackage{amssymb}
 \usepackage{amsmath}
 \usepackage{amsfonts}
 \usepackage{color}
 \usepackage{dcolumn}
 \numberwithin{equation}{section}
 \newcommand{\be}{\begin{equation}}
 \newcommand{\ee}{\end{equation}}
 \newcommand{\bea}{\begin{eqnarray}}
 \newcommand{\eea}{\end{eqnarray}}
 \newcommand{\nn}{\nonumber}
 \newcommand{\rd}{\partial}

\begin{document}

 \begin{titlepage}
  \thispagestyle{empty}

\vspace*{1mm}%
\hfill%
\vbox{
    \halign{#\hfil        \cr
           IPM/P-2010/022
            \cr
          % SUT-P-07-2b   \cr
                     } % end of \halign
      }  % end of \vbox
\vspace*{15mm}%

 % \vspace{2cm}

  \begin{center}
    \font\titlerm=cmr10 scaled\magstep4
    \font\titlei=cmmi10 scaled\magstep4
    \font\titleis=cmmi7 scaled\magstep4

     \centerline{\titlerm  On Born-Infeld Gravity in Three Dimensions}

\vspace*{15mm} \vspace*{1mm} {Mohsen Alishahiha$^{a}$,  Ali
Naseh$^{a,b}$ and Hesam Soltanpanahi$^a$}

 \vspace*{1cm}

{\it ${}^a$ School of physics, Institute for Research in Fundamental Sciences (IPM)\\
P.O. Box 19395-5531, Tehran, Iran \\ }

\vspace*{.4cm}

{\it ${}^b$ Department of Physics, Sharif University of Technology \\
P.O. Box 11365-9161, Tehran, Iran}

\vspace*{2cm}

e-mails: alishah@ipm.ir, naseh@ipm.ir, hsoltan@ipm.ir

\end{center}

\vspace*{2cm}

  \begin{abstract}
In this paper we explore different aspects of three dimensional
Born-Infeld as well as Born-Infeld-Chern-Simons gravity. We show
that the models have AdS and AdS-wave vacuum solutions. Moreover
we observe that although Born-Infeld-Chern-Simons gravity admits a
logarithmic solution, Born-Infeld gravity does not, though it has
a limiting logarithmic solution as we approach the critical point.

  \end{abstract}

\end{titlepage}

 \section{Introduction}

In this paper we would like to study some features of recently
proposed three dimensional gravitational Born-Infeld theory whose
action is given by \cite{GST} \be\label{action1}
I=-\frac{4m^2}{\kappa^2}\int
d^3x\;\sqrt{-\det{g}}\left[\sqrt{-\det\left(1-\frac{1}{m^2}g^{-1}
G\right)}- \left(1+\frac{\Lambda}{2m^2}\right)\right], \ee where
$\kappa^2=16\pi G_3$ and
$G_{\mu\nu}=R_{\mu\nu}-\frac{1}{2}g_{\mu\nu}R$, with $g_{\mu\nu}$
being the three dimensional metric. An interesting feature of this
action is that upon expansion of the action in terms of the
curvature we get cosmological Einstein-Hilbert action at first
order, whereas at second order it leads to NMG action\cite{BHT}.
More interestingly at third order we get the  $R^3$ terms which
have been obtained in \cite{Sinha} by making use of the
requirement that a holographic c-theorem is exist for the theory.
More recently different aspects of Born-Infeld gravity has been
studied in \cite{{Nam:2010dd},{Gullu:2010st}}.

This model together with TMG \cite{{Deser:1981wh},{Deser:1982vy}}
and NMG \cite{BHT} have provided a framework to study three
dimensional gravity with higher curvature terms which have massive
graviton excitations. The hope is that these models can eventually
help us to understand three dimensional quantum gravity. Actually
due to the fact that three dimensional pure gravity with negative
cosmological constant admits BTZ black hole solution
\cite{Banados:1992wn}, it is believed that the theory is
non-trivial quantum mechanically even though at the classical
level it has no propagating degrees of freedom.

It is worth mentioning that adding higher derivative terms to
action will generically lead to instabilities due to the present
of ghost-like modes. Nevertheless it is proposed \cite{Li:2008dq}
that TMG model could be a well defined quantum theory for a proper
boundary condition and at a particular value of the parameters of
the theory. We note, however, that the situation seems to be more
complicated that it was thought at first place. In fact soon after
the proposal \cite{Li:2008dq} it was shown \cite{Grumiller:2008qz}
that the linearized equations of motion of TMG at the critical
point have a new solution.
This now
solution known as logarithmic solution which has the same
asymptotic behavior as AdS wave solution has been first obtained
in \cite{{AyonBeato:2004fq},{AyonBeato:2005qq}}.
Adding this mode will change the nature of the theory (for more
discussions see {\it e.g.}
\cite{{Carlip:2008jk},{Park:2008yy},{Grumiller:2008pr},{Carlip:2008eq},{Carlip:2008qh},
{Giribet:2008bw},{Blagojevic:2008bn},{Li:2008yz}}).

The aim of the present paper is to explore the possibility of
having logarithmic solution in Born-Infeld as well as
Born-Infeld-Chern-Simons gravity. We will also study rotationally
symmetric solutions in the model which could be thought of as
extremal BTZ black holes. By making use of the entropy function
formalism \cite{Sen} we also evaluate the entropy of the
solutions.

The paper is organized as follows. In order to fix our notations
in section two we will review and extend the vacuum solutions of
Born-Infeld gravity where we will also evaluate the entropy of the
extremal black holes of model by making use the entropy function
formalism. In section three we study AdS wave solution in
Born-Infeld gravity where we show that, unlike NMG, the model does
not admits logarithmic solution though it can be reached as a
limiting solution. In section four we extend our study to the
Born-Infeld-Chern-Simons theory whose action is given by
Born-Infeld plus three dimensional gravitational Chern-Simons
term. The last section is devoted to conclusions.

\section{AdS and black hole solutions}

In this section we will review and extend rotationally symmetric
solutions of Born-Infeld theory which includes $AdS_3$ vacuum as
well as solutions with $AdS_2\times S^1$ symmetry. The later
solution may be thought of as extremal black hole solution in the
model.

To find the equations of motion of the model it is useful to
expand the determinant in terms of trace by which the action
\eqref{action1} may be recast to the following form \cite{GST2}
\begin{equation}
I=-\frac{4m^{2}}{\kappa^{2}}\int d^{3}x\,\sqrt{-\det
g}F\left(R,K,S\right),
\end{equation}
where\footnote{In this paper we will only consider the case where
$m>0$ and $\Lambda<0$.}
\begin{equation}
F\left(R,K,S\right)\equiv\sqrt{1+\frac{1}{2m^{2}}\left(R-\frac{1}{m^{2}}K-\frac{1}{12m^{4}}S\right)}-\left(1+\frac{\Lambda}{2m^2}\right),
\end{equation}
with
\begin{equation}
K\equiv R_{\mu\nu}R^{\mu\nu}-\frac{1}{2}R^{2},\qquad
S\equiv8R^{\mu\nu}R_{\mu\alpha}R_{\phantom{\alpha}\nu}^{\alpha}-6RR_{\mu\nu}R^{\mu\nu}+R^{3}.
\end{equation}
%Here $\sigma=\pm 1$ and the above action is reduces to \eqref{action1} for $\sigma=-1$.
 Using this
form of the action the equations of motion read
\cite{Gullu:2010st}
\begin{align}
-\frac{\kappa^{2}}{8m^{2}}T_{\mu\nu} & =-\frac{1}{2}Fg_{\mu\nu}+\left(g_{\mu\nu}\Box-\nabla_{\mu}\nabla_{\nu}\right)F_{R}+F_{R}R_{\mu\nu}\nonumber \\
 & \phantom{=}+\frac{1}{m^{2}}\left\{ 2\nabla_{\alpha}\nabla_{\mu}\left(F_{R}R_{\phantom{\alpha}\nu}^{\alpha}\right)-g_{\mu\nu}\nabla_{\beta}\nabla_{\alpha}\left(F_{R}R^{\alpha\beta}\right)-\Box\left(F_{R}R_{\mu\nu}\right)-2F_{R}R_{\nu}^{\phantom{\nu}\alpha}R_{\mu\alpha}\right.\nonumber \\
 & \phantom{=-\frac{\sigma}{m^{2}}}\left.+g_{\mu\nu}\Box\left(F_{R}R\right)-\nabla_{\mu}\nabla_{\nu}\left(F_{R}R\right)+F_{R}RR_{\mu\nu}\right\} \nonumber \\
 & \phantom{=}-\frac{1}{2m^{4}}\left\{ 4F_{R}R_{\phantom{\rho}\mu}^{\rho}R_{\rho\alpha}R_{\phantom{\alpha}\nu}^{\alpha}+2g_{\mu\nu}\nabla_{\alpha}\nabla_{\beta}\left(F_{R}R^{\beta\rho}R_{\phantom{\alpha}\rho}^{\alpha}\right)+2\Box\left(F_{R}R_{\nu}^{\phantom{\nu}\rho}R_{\mu\rho}\right)\right.\label{eq:Eqn_of_motn}\\
 & \phantom{=-\frac{1}{2m^{4}}}\left.-4\nabla_{\alpha}\nabla_{\mu}\left(F_{R}R_{\nu}^{\phantom{\nu}\rho}R_{\phantom{\alpha}\rho}^{\alpha}\right)+2\nabla_{\alpha}\nabla_{\mu}\left(F_{R}RR_{\phantom{\alpha}\nu}^{\alpha}\right)-g_{\mu\nu}\nabla_{\alpha}\nabla_{\beta}\left(F_{R}RR^{\alpha\beta}\right)\right.\nonumber \\
 & \phantom{=-\frac{1}{2m^{4}}}\left.-\Box\left(F_{R}RR_{\mu\nu}\right)-2F_{R}RR_{\nu}^{\phantom{\nu}\rho}R_{\mu\rho}-g_{\mu\nu}\Box\left(F_{R}R_{\alpha\beta}^{2}\right)+\nabla_{\nu}\nabla_{\mu}\left(F_{R}R_{\alpha\beta}^{2}\right)\right.\nonumber \\
 & \phantom{=-\frac{1}{2m^{4}}}\left.-F_{R}R_{\alpha\beta}^{2}R_{\mu\nu}+\frac{1}{2}g_{\mu\nu}\Box\left(F_{R}R^{2}\right)-\frac{1}{2}\nabla_{\mu}\nabla_{\nu}\left(F_{R}R^{2}\right)+\frac{1}{2}F_{R}R^{2}R_{\mu\nu}\right\}=0 ,\nonumber
\end{align}
where
\begin{equation}
F_{R}=\frac{\partial F}{\partial
R}=\frac{1}{4m^{2}}\left[F+\left(1+\frac{\Lambda}{2m^2}\right)\right]^{-1}.
\end{equation}

In general it is difficult to solve the above equations, though
since we are interested in solutions with $AdS_2\times S^1$
symmetry one may utilize the entropy function formalism
\cite{Sen}. An advantage to work with the entropy function
formalism is that in this formalism not only one can find the
solution but also the entropy of the corresponding solution can be
evaluated.

To proceed we start from an ansatz for the metric with
$AdS_2\times S^1$ symmetry as follows \be\label{ansatz2}
ds^2=v_1\left(-r^2dt^2+\frac{dr^2}{r^2}\right)+v_2(dz+e\,r\,dt)^2.
\ee Then the entropy function is defined by\footnote{If we
compactify the solution into two dimensions the parameter $e$ can
be interpreted as electric field so that $q$ is the electric
charge. Moreover we assume that $e>0$. Since we are considering
the extremal black hole, from dual CFT point of view it means we
are dealing with left moving sector. The right moving sector can
also be studied by assuming $e<0$. On the other hand from three
dimensional point of view the parameter $q$ has to be associated
with the angular momentum of the extremal black hole. Or
equivalently the temperature of the left moving sector may be
given in terms on $1/e$.} \be\label{ef}
 S=2\,\pi(e\,q-f),
\ee where $f=-\frac{8\pi m^{2}}{\kappa^{2}} \sqrt{-\det
g}\;F\left(R,K,S\right)$, evaluated at the ansatz \eqref{ansatz2}.
To find the solution one needs to extremize the entropy function
with respect to the parameters $v_1,v_2$ and $e$.

Using the ansatz \eqref{ansatz2} the entropy function \eqref{ef}
reads
 \be
 f=-\frac{4\pi v_1 v_2^{1/2}}{
 \kappa^2}\left[\frac{1}{4 m v_1^3}
  \sqrt{(- v_2e^2+4m^2v_1^2)^2(3 v_2e^2-4 v_1+4m^2v_1^2)}
  -\left({\Lambda}+2m^2\right)\right].
 \ee
Extremizing the entropy function with respect to  $v_1$, $v_2$ and
$e$ we find three algebraic equations for the
parameters\footnote{To get these equations we assume
$-v_2e^2+4m^2v_1^2>0$. The case of $- v_2e^2+4m^2v_1^2=0$ will be
discussed in the next section.}
 \bea\label{eom}
 E_1&=&
 \left(-3 {v_{{2}}}^{2}{e}^{4}+3 v_{{1}}v_{{2}}{e}^{2}-2 v_{{2}}{v_
 {{1}}}^{2}{e}^{2}{m}^{2}+4 {v_{{1}}}^{3}{m}^{2}-8\,{v_{{1}}}^{4}{m}^{4} \right)\cr
 &&\;\;+2\,m\,{v_1}^3\sqrt{\left(3 v_2e^2-4 v_ 1+4\,{m}^{2}{v_{{1}}}^{2} \right)}
 (\Lambda+2\,m^2)=0,\cr &&\cr
 E_2&=&
 \left(3 {v_2}^{2}{e}^{4}-3 v_1v_2e^2-3 v_2{v_1}^2e^2m^2
 +4 {v_1}^{3}m^2-4\,{v_1}^{4}m^4 \right)\cr
 &&\;\;+m\,{v_1}^3\sqrt{\left(3v_2e^2-4 v_ 1+4\,{m}^{2}{v_{{1}}}^{2} \right)}
 (\Lambda+2\,m^2)=0,\cr &&\cr
 q&=&{\frac {\pi e {v_{{2}} }^{3/2}}{{\kappa}^{2}{v_{{1}}}^{2}{m}} \;\frac{ \left( 9 v_{{2}}{e}^{2}-8 v_{{1}}-4 {m}^{2}{v_{{1}}}^{2}
 \right) }{
 \sqrt { \left(3 v_2e^2-4 v_ 1+4\,{m}^{2}{v_{{1}}}^{2} \right)}}}\,.
 \eea
To solve these equations we note that from the equations $E_1$ and
$E_2$ we find \be E_1-2E_2=(v_2e^2-v_1)(9v_2e^2-4 v_1^2m^2)=0. \ee
Therefore we get two different solutions given by \bea 1)\;
v_2=\frac{v_1}{e^2},\hspace{25mm} 2)\;
v_2=\frac{4m^2{v_1}^2}{9e^2}. \eea In the first case using the
equations of motion for $v_1$ and $e$ one gets \bea
 v_1=-\frac{m^2}{ \Lambda(4m^2+\Lambda)},\;\;\;\;\;\;\;\;\;\;\;\;\;e^2=\frac{\pi}{{\kappa}^{2}\,q\,m}\frac
    {\Lambda+2\,m^2}{\sqrt {-\Lambda(4\,m^2+\Lambda)}}.
\eea Plugging these expressions into the entropy function
\eqref{ef}  we can find the entropy of the corresponding extremal
black hole solution \be S=\frac{4\,\pi^2}{\kappa^{2}\,m\, e}{\frac
    {\Lambda+2\,m^2}{ \sqrt {- {\Lambda }{ \left( 4\,{m}^{2}+\Lambda \right)}}}},
\ee which is physically meaningful for $\Lambda+2\,m^2>0$. As we
will see this is, indeed, the case if we want the theory to
support an AdS vacuum.

From this solution which is essentially a locally $AdS_3$ solution
it is evident that the model admits an $AdS_3$ vacuum solution. In
fact parameterizing the metric of $AdS_3$ solution by \be
ds^2=\frac{l^2}{y^2}(dy^2-2dudv), \ee the radius of the AdS metric
is found to be \be l^2=-\frac{4m^2}{ \Lambda(4m^2+\Lambda)}. \ee
Alternatively one can invert the above equation to find the
cosmological constant in terms of the radius of the $AdS_3$
solution\footnote{Note that from this expression we have
$\Lambda+2m^2=2m^2\sqrt{1-{1}/{m^2l^2}}>0$, as we anticipated.}
\be \Lambda=-2m^2\left(1-\sqrt{1-\frac{1}{m^2l^2}}\right). \ee
Note that to have a real cosmological constant we need to assume
$m^2l^2>1$ (see foot note 3). By making use of this expression it
is interesting to rewrite the entropy of the black hole in terms
of the AdS radius \be S=\frac{\pi^2}{3}\;\frac{1}{2\pi
e}\;\frac{3l}{2G}\sqrt{1-\frac{1}{m^2l^2}}=
2\pi\sqrt{\frac{q}{6}\;\frac{3l}{2G}\sqrt{1-\frac{1}{m^2l^2}}},
\ee which may be compared with the Cardy formula. Indeed using the
c-extremization \cite{KLm} one may identify the central charge of
the dual CFT as follows \cite{{Nam:2010dd},{Gullu:2010st}}
\be\label{c} c_L=\frac{3l}{2G}\sqrt{1-\frac{1}{m^2l^2}}. \ee Using
the Crday formula, \be
S=\frac{\pi^2}{3}T_Lc_L=2\pi\sqrt{\frac{c_L\bar{L}_0}{6}}, \ee we
are led to the following identification \be
\bar{L}_0=q,\;\;\;\;\;\;\;\;\;\;\;\;T_L=\frac{2}{\pi}\sqrt{\frac{Gq}{l\sqrt{1-\frac{1}{m^2l^2}}}}.
\ee The identification of the temperature may also be understood
from the period of the compact direction in the ansatz
\eqref{ansatz2}.

On the other hand for the second solution one finds \bea
v_1&=&{\frac
{9}{8}}\,{\frac{188\,{m}^{4}-324\,\Lambda\,{m}^{2}-81\,{
\Lambda}^{2}\pm 9\,\sqrt{\left(9\,\Lambda+ 34\,{m}^{2} \right)
\left( 9\,\Lambda+2\,{m}^{2} \right)}  \left( \Lambda+2\,{m}^{2}
\right)}{{m}^{2} \left( 52\,{m}^{4}-
972\,\Lambda\,{m}^{2}-243\,{\Lambda}^{2}\right)}},\cr &&\cr
e^2&=&-\frac{32\pi}{27\,\kappa^2\,q}\frac{m^2\,{v_1}^{2}}{\sqrt{\frac{4}{3}\,{v_1}^2{m}^{2}-\,v_1}}
\eea that requires to have  $9\Lambda+2\,{m}^{2}>0$ which is
stronger condition than we had in the previous case. Actually this
solution may be interpreted as a warped black hole solution. It is
easy to find the entropy of the corresponding black hole using the
entropy function \eqref{ef} \bea
 S_2=\frac{16\pi^2m\,{v_1}^{2}}{81\,\kappa^2\,e\sqrt{\frac{4}{3}\,{v_1}^2{m}^{2}-\,v_1}}
 \left(64\,m^3v_1-60\,m-27\,(\Lambda+2\,m^2)\sqrt{\frac{4}{3}\,{v_1}^2{m}^{2}-\,v_1}\right)
 \eea

 \section{AdS wave solution and Log gravity}

Having found the AdS vacuum solution, it is interesting to study
AdS wave solution for Born-Infeld gravity. AdS wave solutions for
TMG and NMG models have been studied in
\cite{{AyonBeato:2004fq},{AyonBeato:2005qq},{AyonBeato:2009yq}}
where it was shown that at the critical value of the parameters
the solution develops logarithmic behaviors. The same situation
has also been obtained in bi-gravity \cite{Afshar:2009rg}. In this
section we will see that the situation is a little bit different
for Born-Infeld gravity.

As we have seen in the previous section the model admits an
$AdS_3$ vacuum solution which can be parametrized as
\be\label{AdS3}
ds^2=\frac{l^2}{y^2}\left(dy^2-2dudv\right),\;\;\;\;{\rm
with}\;\;\;\;l^2=-\frac{4m^2}{\Lambda(\Lambda+4m^2)}. \ee To
proceed, based on this notation, we consider an ansatz for the AdS
wave solution as follows
 \bea\label{wave}
 ds^2=\frac{l^2}{ y^2}\left[dy^2-2dvdu-G(u,y)du^2\right],
 \eea
where $G(u,y)$ is an arbitrary function to be determined by
equations of motion. Plugging this ansatz in the equations of
motion \eqref{eom} one finds that the radius, $l$, has the same
expression as that for the AdS solution. Moreover the function $G$
obeys the following  differential equation
 \bea
  \frac{y^4\frac{\rd^4G}{\rd y^4}
 +2\,y^3\frac{\rd^3G}{\rd y}-\,m^2\,l^2\,\left(y^2\frac{\rd^2G}{\rd y^2}
 -y\frac{\rd G}{\rd y}\right)}{y^2\,l\,m\,\sqrt{m^2\,l^2-1}}=0.\label{eom2}
 \eea
One might naively think that the dominator of the above equation
is redundant. Actually this is not the case. In fact the dominator
plays a crucial rule especially at the critical point $m^2l^2=1$.
Moreover for large $ml$ limit one expects that the above equation
reduces to that in NMG model. Indeed evaluating the limit
correctly we find that there are contributions from the dominator
too. More precisely at large $ml$ limit one finds \bea
  \frac{1}{y^2\,m^2l^2}\Bigg[y^4\frac{\rd^4G}{\rd y^4}
 +2\,y^3\frac{\rd^3G}{\rd y}-\,\frac{2m^2\,l^2+1}{2}\,\left(y^2\frac{\rd^2G}{\rd y^2}
 -y\frac{\rd G}{\rd y}\right)\Bigg]+{\cal O}(\frac{1}{m^4l^4})=0,
\eea which at leading order is exactly the equation we have in NMG
model \cite{AyonBeato:2009yq}.

The standard solution for the above fourth order Euler-Fuchs
differential equation is
 $G=y^\alpha$ where $\alpha$ satisfies the following equation
 \bea
 \alpha(\alpha-2)\left[(\alpha-1)^2-\,m^2l^2\right]=0.\label{alpha}
 \eea
 Thus the generic solution for AdS-wave is
 \bea
 G(u,y)=G_0(u)+G_2(u)y^2+G_+(u)\left(\frac{y}{ l}\right)^{1+m\,l}+G_-(u)\left(\frac{y}{ l}\right)^{1-m\,l}
 \eea
 where $G$'s are arbitrary functions of retarded time $u$. Note that the first two terms can be
eliminated by a coordinate transformation \cite{AyonBeato:2005qq}.

It is natural to look for a possibility of having multiplicities
in the roots of the characteristic equation \eqref{alpha}. In fact
we see that the equation has a multiplicity in the roots for
$m^2l^2=1$. It is, however, important to note that at this point
the dominator of the equation \eqref{eom2} is zero. Therefore we
are not allowed to set $m^2l^2=1$. Nevertheless it is natural to
look for a possible limiting solution when $m^2l^2\rightarrow 1$.
Indeed starting from the following ansatz
  \bea
 G(u,y)=\ln\left(\frac{y}{l}\right)\left[G_1(u)\left(\frac{y}{
 l}\right)^2+G_2(u)\right],
 \eea
and plugging it into the equation \eqref{eom2}, one arrives at \be
\frac{2(G_1l^2-G_2 y^2)}{y^2\, l^3\, m}\sqrt{-1+m^2l^2}, \ee which
is zero in the limit of $m^2l^2\rightarrow 1$. It is worth noting
that although the limit is well defined the log solution is not a
solution of the equations of motion and indeed it can be treated
as a limiting solution. This is in contrast with what happens in
TMG and NMG where the solution at the critical point is a solution
of the equations of motion.

%-------------------------------------------------------------------------------------------

 \section{Adding Chern-Simons term}

In this section we would like to extend the Born-Infeld gravity by
adding a three dimensional gravitational Chern-Simons term to the
Born-Infeld action. The gravitational Chern-Simons action is given
by
 \be
 I_{CS}=\frac{1}{2\kappa^2\mu}\int\,d^3x\,\sqrt{-g}\epsilon^{\lambda\mu\nu}
 \left(\Gamma^\rho_{~\lambda\sigma}\rd_\mu\Gamma^\sigma_{~\rho\nu}
 +\frac{2}{3}\Gamma^\rho_{~\lambda\sigma}\Gamma^\sigma_{~\mu\tau}\Gamma^\tau_{~\nu\rho}\right).\label{CS}
 \ee
Adding this term to the action the equations of motion \eqref{eom}
will be corrected by the term $\frac{1}{\mu}C_{\mu\nu}$, with
$C_{\mu\nu}$ being the Cotton tensor. On the other hand since the
Cotton tensor is zero for and $AdS_3$ solution, the model still
admits an $AdS_3$ vacuum solution which is indeed the same
solution as we had in the previous section \eqref{AdS3}.
%\be
%ds^2=\frac{l^2}{y^2}\left(dy^2-2dudv\right),\;\;\;\;{\rm with}\;\;\;\;l^2=-\frac{4m^2}{\Lambda(\Lambda+4m^2)}.
%\ee
It is then obvious that the black hole solutions of Born-Infeld
gravity which are locally AdS, are still solutions of the modified
model. Nevertheless since the modified model is a parity violating
model the nature of the theory changes drastically. In particular
the central charges of left and right moving sectors of the dual
CFT are not equal.

To find the corresponding central charges using the
c-extremization procedure, we recall that when the left and right
central charges are not equal what the c-extremization procedure
computes is the mean value of the central charges \cite{KLm}. In
other words applying the procedure for our case the result of the
previous section \eqref{c} should be read as follows
\be\label{CSc+} \frac{c_L+c_R}{2}=\frac{3l}{2G}\sqrt{1-\frac{1}{
m^2\,l^2}}. \ee On the other hand due to the present of the
Chern-Simons term the whole action is diffeomorphism invariant up
to a boundary term. From boundary theory point of view this shows
itself in the gravitational anomaly in the dual CFT. Therefore the
difference between the central charges is non-zero and in fact
should be proportional to the coefficient of the Chern-Simons
term. More precisely we have \cite{Kraus:2005zm} \be
 c_L-c_R=-\frac{3}{ G\mu}.\label{CSc-}
 \ee
Therefore from the equations \eqref{CSc+} and \eqref{CSc-} we get
 \bea
c_L=\frac{3l}{2G}\left(\sqrt{1-\frac{1}{ m^2\,l^2}}-\frac{1}{\mu
l}\right),\;\;\;\;\;\;\;\;\;\;\;\;
c_R=\frac{3l}{2G}\left(\sqrt{1-\frac{1}{ m^2\,l^2}}+\frac{1}{\mu
l}\right).
 \eea
Following the suggestion of \cite{Li:2008dq} it is natural to see
if there is any point in the moduli space of the parameters where
the theory could be chiral. In fact looking that the central
charge we observe that there is a possibility to set $c_L=0$ which
could lead to a consistent chiral theory with a proper boundary
condition. Actually in this case we find a chiral line along which
the left handed central charge vanishes\footnote{If we allow $\mu$
to be also negative $c_R$ can be set to zero too.}
\be\label{chiral} \sqrt{1-\frac{1}{ m^2\,l^2}}=\frac{1}{\mu l}.
\ee Of course it is clear that for large $m^2l^2$ limit the above
expression reduces to that in NMG model.

Having had AdS vacuum it is worth looking for AdS wave in this
model. In fact starting from the AdS wave ansatz \eqref{wave} the
equations of motion reduces to the following differential equation
for function $G$
 \bea
 \frac{y^4\frac{\rd^4G}{\rd y^4}
 +\left(2-\frac{m}{\mu}\sqrt{{m^2\,l^2}-1}\right)\,y^3\frac{\rd^3G}{\rd y}
 -m^2\,l^2\,\left(y^2\frac{\rd^2G}{\rd y^2}
 -y\frac{\rd G}{\rd y}\right)}{y^2\,l\,m\,\sqrt{{m^2\,l^2}-1}}=0.
 \eea
In large $m^2l^2$ limit the above equation may be expanded to get
\bea
  \frac{1}{y^2\,m^2l^2}\Bigg[y^4\frac{\rd^4G}{\rd y^4}
 +\left(2-\frac{l m^2}{\mu}\right)y^3\frac{\rd^3G}{\rd y}-\,\frac{2m^2\,l^2+1}{2}\,\left(y^2\frac{\rd^2G}{\rd y^2}
 -y\frac{\rd G}{\rd y}\right)\Bigg]+{\cal O}(\frac{1}{m^4l^4})=0,
\eea which at leading order coincides with that for generalized
massive gravity \cite{AyonBeato:2009yq}. Note that to get the
correct expansion it was crucial to take into account the
contributions of the dominator.

It is clear that the most general solution of the above
differential equation is in the form of $y^\alpha$ with constant
$\alpha$ satisfying the following characteristic equation

 \be\label{ch2}
 \alpha(\alpha-2)\left[(\alpha-1)^2-(\alpha-1)\frac{m}{\mu}\sqrt{m^2\,l^2-1}-m^2\,l^2\right]=0.
 \ee
Therefore the generic solution is
 \bea
 G(u,y)&=&
 G_+(u)\left(\frac{y}{ l}\right)^{\frac{1}{2\mu}(\, 2\mu+m\sqrt{m^2l^2-1}+\sqrt{m^2(m^2l^2-1)
 +4\mu^2m^2l^2}\,)}\nn\\
 &+&G_-(u)\left(\frac{y}{ l}\right)^{\frac{1}{2\mu}(\, 2\mu+m\sqrt{m^2l^2-1}-\sqrt{m^2(m^2l^2-1)
 +4\mu^2m^2l^2}\,)}.
 \eea
To write the above solution we used the fact that the quadratic
dependence of the solution can be eliminated by a coordinate
transformation.

It is interesting to note that at the critical chiral line given
by the equation \eqref{chiral} where $c_L=0$ the characteristic
equation degenerates leading to a new logarithmic solution \be
G(u,y)=G_1(u)\ln\left(\frac{y}{ l}\right)+G_2(u)\left(\frac{y}{
l}\right)^{
 m^2l^2+1}.
\ee We observe that although the Born-Infeld gravity does not
support the log gravity solution adding Chern-Simons extends the
range of parameters to accommodate a log gravity solution. We note
also that although at $ml=1$ the characteristic equation
\eqref{ch2} has multiplicity in the roots, the model is only well
defined for $ml>1$ and we are not allowed to set $ml=1$. Moreover,
unlike  the case considered in previous section, in the present
case the limit of $ml\rightarrow 1$ is not a well defined too.
This means that we do not have a well defined limiting logarithmic
solution as well.

\section{Conclusions}

In this paper we have studied some aspects of recently proposed
Born-Infeld gravity. A nice feature of the proposed action is that
upon expanding the action in powers of the curvature at second
order the theory reduces to NMG model and at third order the $R^3$
terms coincident with terms obtained by making use of the
holographic c-theorem \cite{Sinha}. Indeed the action was
constructed to have such a feature. It is important to note that
to get Born-Infeld action one has to sum up infinite terms.
Truncating the infinite sum to a summation of finite terms would
drastically change the physical content of the
theory\footnote{Actually such an effect has been led to an
interesting inflationary model
\cite{{Silverstein:2003hf},{Alishahiha:2004eh}}.}.

In particular NMG model has a critical point given by $m^2l^2=1/2$
where the central charge of the dual CFT theory is zero and the
model admits a new vacuum solution that is not asymptotically
locally $AdS_3$ \cite{AyonBeato:2009yq}. It is believed that at
this point the theory is dual to a LCFT
\cite{{Grumiller:2009sn},{Alishahiha:2010bw}}. Nevertheless when
we are considering the Born-Infeld gravity there is noting special
at $m^2l^2=1/2$. In other words summing up the infinite terms,
somehow, resolves this point. On the other hand one might suspect
that the effect of the infinite summation is just shifting the
value of the critical point to $m^2l^2=1$. We note, however, that
this is not the case.  Indeed we encounter an interesting effect
in Born-Infeld gravity.

Actually the critical point where the central charge vanishes is a
singular point in the moduli space of the parameters of the
theory. In other words the point can only be reached in the limit
of $m^2l^2\rightarrow 1$. Taking into account that usually we get
a logarithmic solution at the critical point, this means that the
log gravity is not a solution of the model, though we can have it
as a limiting solution. It would be interesting to explore the
meaning of this effect from dual CFT description.

We have also seen that adding Chern-Simons terms to the
Born-Infeld gravity the situation will change  drastically.
Although the $ml=1$ point is still a singular point in the moduli
space of the parameters, there is a critical chiral line where the
left handed central charge vanishes and the solution develops a
logarithmic term. Form our experience of TMG and NMG model
\cite{{Skenderis:2009nt},{Grumiller:2009mw},{Grumiller:2009sn},{Alishahiha:2010bw}}
we would expect that the theory along this critical line is dual
to a LCFT. It would  also be interesting to understand this
correspondence better.

In this paper we have studied Born-Infeld gravity with the
assumption $m>0, \Lambda<0$. It is possible to relax this
assumption to the cases where $m<0$ or $m^2<0$. In these case it
is possible to have other critical points. Moreover throughout our
paper we have chosen a specific sign for the action. It is,
however, possible to use another signs by modifying the action
with some parameters, named by $\eta$ and $\sigma$ in
\cite{Nam:2010dd}. Actually we have done all of our computations
for generic $\eta$ and $\sigma$. But we noticed that the most
physical choice is $\eta=1$ and $\sigma=-1$. Therefore we have
presented our results only for this specific choice.

We note that the logarithmic solutions we have found in this paper
provide gravity descriptions for logarithmic CFT's which might
have application in condensed matter physics.

\section*{Acknowledgments}

We would like to thank R. Fareghbal and A. Mosaffa  for useful
discussions.
We also would like to thank D. Grumiller for a comment on the first version of the manuscript.

\end{document}